\documentclass[prl,twocolumn,showpacs,superscriptaddress]{revtex4}
\usepackage{graphicx}
\usepackage{amsfonts,amsmath,amssymb}

\begin{document}
\title{Quantum-control approach to realizing a Toffoli gate in circuit QED}
\author{Vladimir M. Stojanovi\'c}
\affiliation{Department of Physics, University of Basel,
 Klingelbergstrasse 82, CH-4056 Basel, Switzerland} 

\author{A. Fedorov}
\affiliation{Department of Physics, ETH Z\"urich, CH-8093 Z\"urich, Switzerland}

\author{A. Wallraff}
\affiliation{Department of Physics, ETH Z\"urich, CH-8093 Z\"urich, Switzerland}

\author{C. Bruder}
\affiliation{Department of Physics, University of Basel,
 Klingelbergstrasse 82, CH-4056 Basel, Switzerland} 

\date{\today}
\begin{abstract}
We study the realization of a Toffoli gate with superconducting qubits
in a circuit-QED setup using quantum-control methods.  Starting with
optimized piecewise-constant control fields acting on all qubits and
typical strengths of $XY$-type coupling between the qubits, we
demonstrate that the optimal gate fidelities are affected only
slightly by a ``low-pass'' filtering of these fields with the typical
cutoff frequencies of microwave driving.  Restricting ourselves to the
range of control-field amplitudes for which the leakage to the
non-computational states of a physical qubit is heavily suppressed, we
theoretically predict that in the absence of decoherence and leakage, within
$75$\:ns a Toffoli gate can be realized with intrinsic fidelities
higher than $90\%$, while fidelities above $99\%$ can be
reached in about $140$\:ns.
\end{abstract}
\maketitle 

Superconducting (SC) qubits~\cite{SCqubitReviews} have come a long way
since the realization that Josephson physics in SC circuits can be
utilized to prepare well-defined few-level quantum
systems~\cite{Martinis+:85}.  Coupling such qubits is essential for
quantum computation. The most successful approaches up to now rely on
coupling all the qubits in an array to an ``interaction bus,'' a
central coupling element having the form of a transmission line with
electromagnetic modes. In the circuit quantum electrodynamics (circuit
QED) regime strong coupling between the qubits and the confined
photons is realized~\cite{Blais++:04,Wallraff++:04,Majer++:07}; the
cavity photons induce a long-range coupling between the qubits.  The
combination of transmon qubits~\cite{Koch++:07} and co-planar
microwave cavities represents the state-of-the-art of microwave
quantum optics.

The environmental degrees of freedom limit the time over which quantum
coherence can be preserved. While they are similar to charge 
qubits, transmons have a much larger total capacitance such that the charging
energy is significantly smaller than the Josephson energy. A small charge
dispersion of the energy eigenstates leads to a reduced sensitivity to
charge noise and longer dephasing times ($T_{2}$)~\cite{Koch++:07}. 
Recently, a significant progress has been achieved~\cite{Kim++:11,Paik++:11}, 
with $T_{2}$ times being increased by an order of magnitude from 
$T_{2}\sim 1\:\mu s$ to $T_{2}\sim 20\:\mu s$.

Given that two-qubit gates with SC qubits have been demonstrated with
fidelities higher than $90\%$~\cite{DiCarlo++:09}, a key challenge now
is to realize three-qubit ones with shortest possible gate times. An
example is the Toffoli gate (controlled-controlled-NOT), which is
relevant for quantum-error correction~\cite{MerminBook} and has
already been implemented with trapped ions~\cite{Monz++:09} and
photonic systems~\cite{Lanyon++:09} with respective fidelities of
$71\%$ and $81\%$. Very recently, several groups realized a Toffoli gate using
superconducting
circuits~\cite{Mariantoni++:11,Fedorov++:11,Reed++:11}. The Toffoli gate was
implemented by a sequence of single- and two-qubit gates (direct approach).

\begin{figure}
\includegraphics[width=0.485\textwidth]{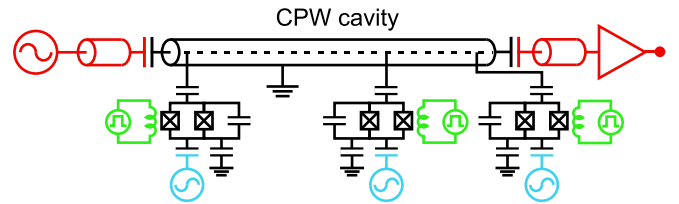}
\caption{\label{CircuitQEDsketch} Lumped-element circuit diagram of
  three transmon qubits coupled to a superconducting transmission-line
  resonator. The resonator serves as a coupling bus for the qubits and
  is also used for the readout of their states (red). Local flux lines
  (blue) allow for individual control of the qubit frequencies on the
  nanosecond time-scale.  Microwave lines (green) are used to create
  control fields, each acting on its corresponding qubit.}
\end{figure}

In this paper, we adopt an alternative approach and study the realization of
a Toffoli gate with SC qubits in a circuit-QED setup (for an
illustration, see Fig.~\ref{CircuitQEDsketch}) by applying
quantum-control methods~\cite{D'AlessandroBook} to the effective
$XY$-type Hamiltonian of an interacting three-qubit array. We do so
for realistic qubit-qubit coupling strengths and under typical
experimental constraints on the qubit decoherence times.

We first determine optimal piecewise-constant control fields and
evaluate the resulting gate fidelities. Then we discuss how these
fidelities are affected when the control pulses are smoothened by
eliminating their high-frequency Fourier components (spectral
``low-pass'' filtering). Our most important theoretical 
prediction is that within
$75$\:ns a Toffoli gate can be realized with intrinsic fidelities
(in the absence of decoherence and leakage) higher than $90\%$, while 
fidelities higher than $99\%$ can be obtained in 
approximately $140$\:ns.

Methods of quantum control~\cite{D'AlessandroBook} have been put
forward in a number of theoretical proposals for realizing quantum
logic gates with SC qubits~\cite{SCqubitControl,Motzoi++:09,Gambetta++:11}, thus
complementing studies that solely involve time-independent
Hamiltonians~\cite{GaliautdinovGeller}. More fundamentally, recent
studies in operator (state independent) control~\cite{Zhang+Whaley:05}
have been focusing on interacting systems and employing the concept
of local control~\cite{LocalControl,Heule++}. The essential idea is
that systems such as coupled spin-$1/2$ chains, models of interacting
qubit arrays, can often be controlled by acting on a small
subsystem. For instance, controlling only one end spin of an
$XXZ$-Heisenberg chain ensures complete controllability of the
chain~\cite{LocalControl,Heule++}. Yet, in view of the currently
available few-qubit experimental setups~\cite{Filipp++:11}, more
important than restricting control to only one qubit is to be able to
carry out an optimal control-pulse sequence within times much shorter
than $T_{2}$. This is necessary in order to minimize the undesired
decoherence effects~\cite{SCqubitReviews}.

Under the condition of resonant driving and assuming that
the qubits are in resonance with one another, the effective
(time-independent) $XY$-type (flip-flop) qubit-qubit interaction
Hamiltonian is given by
\begin{equation}\label{H_0a}
H_{0}=\sum_{i< j}J_{ij}(\sigma_{ix}\sigma_{jx}
+\sigma_{iy}\sigma_{jy})\:,
\end{equation}
where $\sigma_{ix}$, $\sigma_{iy}$, and $\sigma_{iz}$ 
are the Pauli matrices. The system is acted upon 
by time-dependent Zeeman-like control fields described by the Hamiltonian  
\begin{equation}
 H_c(t)=\sum_{i=1}^{3}\left[\Omega^{(i)}_{x}(t)\:\sigma_{ix}+
\Omega^{(i)}_{y}(t)\:\sigma_{iy}\right]\:,
\end{equation}
thus the system dynamics is governed by the total Hamiltonian
$H(t)=H_{0}+ H_c(t)$.

The control fields $\Omega^{(i)}_{x}(t)$ and $\Omega^{(i)}_{y}(t)$ can
be implemented using arbitrary wave generators (recall
Fig.~\ref{CircuitQEDsketch}), which can produce an arbitrary signal
with frequencies up to $500$\:MHz with minor distortions. 
In qubits based on weakly-anharmonic
oscillators, leakage from the two-dimensional qubit Hilbert space
(computational states) is the leading source of errors at short gate
times~\cite{Fazio+:99,Gambetta++:11}.  This is especially pronounced
if the control bandwidth is comparable to the anharmonicity~\cite{Motzoi++:09}. 
We therefore impose an additional constraint on the control fields,
viz. the condition that
\begin{equation}\label{constraint}
\Omega_{\textrm{max}}=\textrm{max}_{i,t}
\sqrt{[\Omega^{(i)}_{x}(t)]^{2}+[\Omega^{(i)}_{y}(t)]^{2}} 
\end{equation}
is smaller than some threshold value for the transmon to be a
well-defined two-level system.  
For typical anharmonicities of transmon qubits ($300-400$\:MHz) and
values of $\Omega_{\textrm{max}}$ of $100-130$\:MHz the error
associated with this leakage should not exceed a few percent.  While
control schemes explicitly involving the higher levels of a physical
qubit are in principle conceivable, in the present study we aim for
simplicity and will in the following determine the control pulses for
a system of qubits that are genuine two-level systems.

Before evaluating numerically the optimal control pulses we would like
to comment on the controllability aspects of the problem.
Using the standard algorithm (see, for example,
Ref.~\onlinecite{D'AlessandroBook}), it is straightforward to check
that the dynamical Lie algebra of the system, generated by the
skew-Hermitian operators $-iH_{0}$, $-i\sigma_{ix}$, and
$-i\sigma_{iy}$ ($i=1,2,3$), has dimension $63=d^{2}-1$ ($d=8$ is the
dimension of the Hilbert space of the system). This Lie algebra 
is isomorphic to $su(d=8)$ and the system is completely
(operator) controllable. An arbitrary quantum gate can thus in
principle be realized using properly designed control fields.

Our goal is to find the time dependence of control fields
$\Omega^{(i)}_{x}(t)$ and $\Omega^{(i)}_{y}(t)$ for realizing a quantum
Toffoli gate. We start our analysis with simple piecewise-constant
control fields~\cite{Heule++} acting on all three qubits in
alternation in the $x$- and $y$ directions with control amplitudes
$\Omega^{(i)}_{x,n}$ and $\Omega^{(i)}_{y,n}$
($n=1,\ldots,N_{t}/2\:;\:i=1,2,3$).

At $t=0$ control pulses are applied in the $x$ direction to all three
qubits with constant amplitudes $\Omega^{(i)}_{x,1}$ during the time
interval $0\leq t\leq T$. The Hamiltonian of the system is then
$H_{x,1}\equiv H_0+\sum_{i=1}^{3}\:\Omega^{(i)}_{x,1}\sigma_{ix}$.
Then $y$ control pulses with amplitudes $\Omega^{(i)}_{y,1}$ are
applied during the interval $T\leq t\leq 2T$, whereby the system
dynamics is governed by $H_{y,1}\equiv
H_0+\sum_{i=1}^{3}\:\Omega^{(i)}_{y,1}\sigma_{iy}$.  This sequence of
alternating $x$ and $y$ control pulses is repeated until $N_t$ pulses
have been completed at the gate time $t_g\equiv N_tT$. The
time-evolution operator $U(t=t_g)$ is then obtained as a product of
the consecutive $U_{x,n}\equiv\exp(-iH_{x,n}T)$ and
$U_{y,n}\equiv\exp(-iH_{y,n}T)$, where $n=1,\ldots,N_{t}/2$.

For varying choices of $N_t$ and $T$, the $3N_t$ control amplitudes
are determined so as to maximize the fidelity
\begin{equation}\label{deffidelity}
F(t_g)=\frac{1}{8}\big|\mathrm{tr}
\big[U^{\dag}(t_g)U_{\mathrm{TOFF}}\big]\big|\:,
\end{equation}
where $U_{\mathrm{TOFF}}$ is the Toffoli gate. The numerical
maximization over these amplitudes is carried out using the
Broyden-Fletcher-Goldfarb-Shanno (BFGS)
algorithm~\cite{NRfortranBook}, a standard second-order
quasi-Newton-type procedure. Based on an initial guess for the control
amplitudes, the algorithm generates iteratively new sequences of
amplitudes such that in each iteration step the fidelity is increased,
terminating when the desired accuracy is reached. This procedure is
repeated for multiple ($\sim 200$) initial guesses to avoid getting
trapped in local (instead of the global) maxima of $F(t_g)$. The
optimization is performed under the constraint 
$\Omega_{\textrm{max}}<130$\:MHz.

While piecewise-constant control pulses are convenient
as a starting point for a theoretical analysis, the actual
pulse-shaping hardware cannot generate such fields with arbitrarily-high
frequency components. Turning the time course of piecewise-constant 
control pulses into an optimized shape can be considered the central 
problem of numerical optimal control~\cite{Machnes++:11}. 

We therefore perform spectral filtering of our optimal control fields.
Quite generally, after acting with a frequency-filter function $f(\omega)$ 
on the Fourier transforms $\mathcal{F}[\Omega^{(i)}_j(t)]$ of the 
optimal fields $\Omega^{(i)}_j(t)$ ($j=x,y$), one switches back to the time
domain via inverse Fourier transformation to obtain the filtered fields
$\widetilde{\Omega}_j(t)$:
\begin{equation}\label{filtering_procedure}
\widetilde{\Omega}^{(i)}_j(t)=\mathcal{F}^{-1}\big[f(\omega)
\mathcal{F}[\Omega^{(i)}_j(t)]\big] \quad (j=x,y)\:.
\end{equation}
In particular, we consider an {\em ideal low-pass} filter, which 
removes frequencies above the cut-off $\omega_0$ and below $-\omega_0$.
In other words, $f(\omega)=\theta(\omega+\omega_0)-\theta(\omega-\omega_0)$\:,
where $\theta(x)$ is the Heaviside function. When applied to our 
piecewise-constant control fields, the transformations in 
Eq.~\eqref{filtering_procedure} can be carried out semi-analytically.
They lead to 
\begin{equation}\label{tildeh}
\begin{split}
\widetilde{\Omega}^{(i)}_x(t)=\frac{1}{\pi}\sum_{n=1}^{N_t/2}
\Omega^{(i)}_{x,n}\big[a_{2n-1}(t)-a_{2n-2}(t)\big]\:,\\
\widetilde{\Omega}^{(i)}_y(t)=\frac{1}{\pi}\sum_{n=1}^{N_t/2}
\Omega^{(i)}_{y,n}\big[a_{2n}(t)-a_{2n-1}(t)\big]\:,
\end{split}
\end{equation}
where $a_{m}(t)\equiv\mathrm{Si}\big[\omega_0(mT-t)\big] \quad
(m\in\mathbb{N})$ and $\mathrm{Si}(x)\equiv\int_0^{x}(\sin{t}/t)dt$
stands for the sine integral. Based on Eq.~\eqref{tildeh}, we
numerically determine the time-evolution operators corresponding to
the filtered control fields using a product-formula approach (for
details, see the Appendix in Ref.~\onlinecite{Heule++}).  We then
obtain the fidelities $\widetilde{F}(t_g)$ corresponding to the
filtered fields from an analog of Eq.~\eqref{deffidelity}.
\begin{table}
\caption{\label{TableOfFidelities} Examples of calculated intrinsic
  Toffoli-gate fidelities for optimal piecewise-constant control
  fields ($F$) and their low-pass filtered versions ($\widetilde{F}$),
  both corresponding to gate times $t_{g}$ given in units of $J^{-1}$
  where $J=30$\:MHz.  The two different values shown for
  $\widetilde{F}$ and $\widetilde{\Omega}_{\textrm{max}}$ correspond
  to respective high-frequency cutoffs of $\omega_{0}=500$\:MHz and
  $\omega_{0}=450$\:MHz (in brackets).  }
\begin{ruledtabular}
\begin{tabular}{r c c c c}
$N_{t}$ & $t_{g}$\:[ns] & $F$\:[\%] & $\widetilde{F}$\:[\%] &
$\widetilde{\Omega}_{\textrm{max}}$\:[MHz]\\ \hline
14 & 75.0 &  92.92 & 92.08 (91.43) & 102.7 (96.0) \\
12 & 76.0 &  91.74 & 91.38 (91.20) & 96.5 (96.2) \\
10 & 81.3 &  91.91 & 91.39 (91.35) & 107.5 (104.7) \\
20 & 139.2 & 99.72 & 99.29 (99.28) & 111.2 (112.2) \\
18 & 165.0 & 99.72 & 99.45 (99.35) & 102.9 (94.9) \\
16 & 180.0 & 99.00 & 98.79 (98.77) & 119.1 (116.0) \\
18 & 180.0 & 99.70 & 99.52 (99.40) & 116.4 (107.2) \\
30 & 195.0 & 99.99 & 99.57 (99.15) & 94.8 (83.9) \\
28 & 198.9 & 99.99 & 99.23 (98.68) & 126.4 (119.0) \\
18 & 215.0 & 99.78 & 99.61 (99.59) & 102.7 (100.8) \\
20 & 213.3 & 99.96 & 99.70 (99.70) & 105.8 (102.0) \\
24 & 205.0 & 99.99 & 99.84 (99.72) & 129.0 (122.9) \\
22 & 207.2 & 99.98 & 99.88 (99.78) & 118.7 (113.4) \\
22 & 210.5 & 99.99 & 99.89 (99.86) & 119.8 (114.7)\\
24 & 215.0 & 99.99 & 99.61 (99.47) & 129.6 (126.9)\\
22 & 224.6 & 99.99 & 99.91 (99.79) & 108.9 (101.0)\\
22 & 230.0 & 99.99 & 99.96 (99.93) & 126.6 (120.4)\\
\end{tabular}
\end{ruledtabular}
\end{table} 

Our numerical results are summarized in Table~\ref{TableOfFidelities}.
It shows examples of calculated Toffoli-gate fidelities for optimal
piecewise-constant control fields ($F$) and their low-pass filtered
versions ($\widetilde{F}$), for two different high-frequency cutoffs
($\omega_{0}=500$\:MHz and $\omega_{0}=450$\:MHz). The last column of
the table shows the maximum $\widetilde{\Omega}_{\textrm{max}}$ of
$\sqrt{[\tilde{\Omega}^{(i)}_{x}(t)]^{2}+[\tilde{\Omega}^{(i)}_{y}(t)]^{2}}$
over all qubits and all times and obeys the constraint discussed after
Eq.~\eqref{constraint}. The fidelities corresponding to the
piecewise-constant fields are virtually unaffected by the filtering
process, since the highest frequencies achievable by current
pulse-shaping hardware are much larger than the qubit-qubit coupling
strengths.

\begin{figure}
\includegraphics[width = 0.425\linewidth,clip=true]{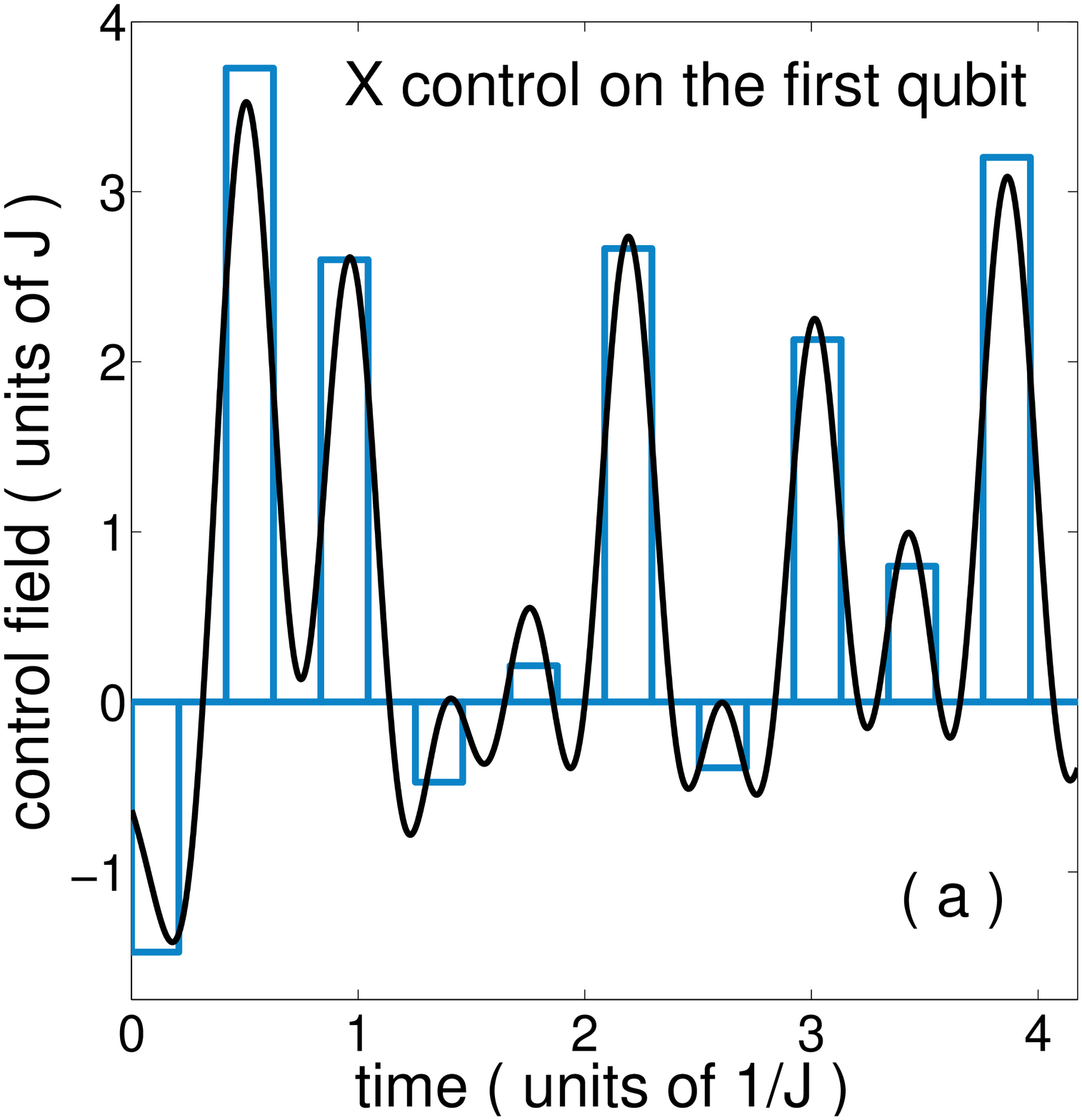}%
\quad\includegraphics[width = 0.425\linewidth,clip=true]{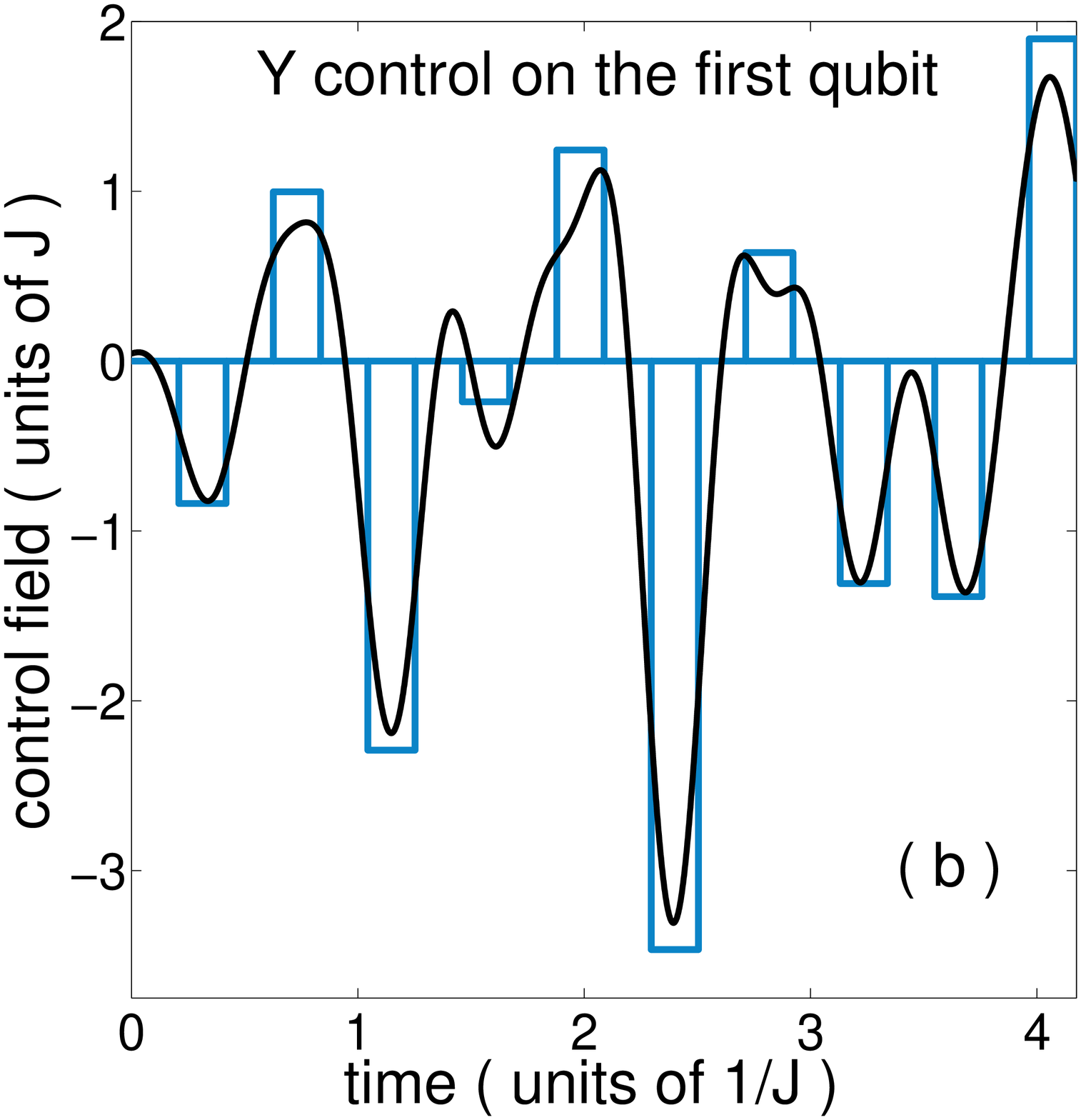}\\
\includegraphics[width = 0.425\linewidth,clip=true]{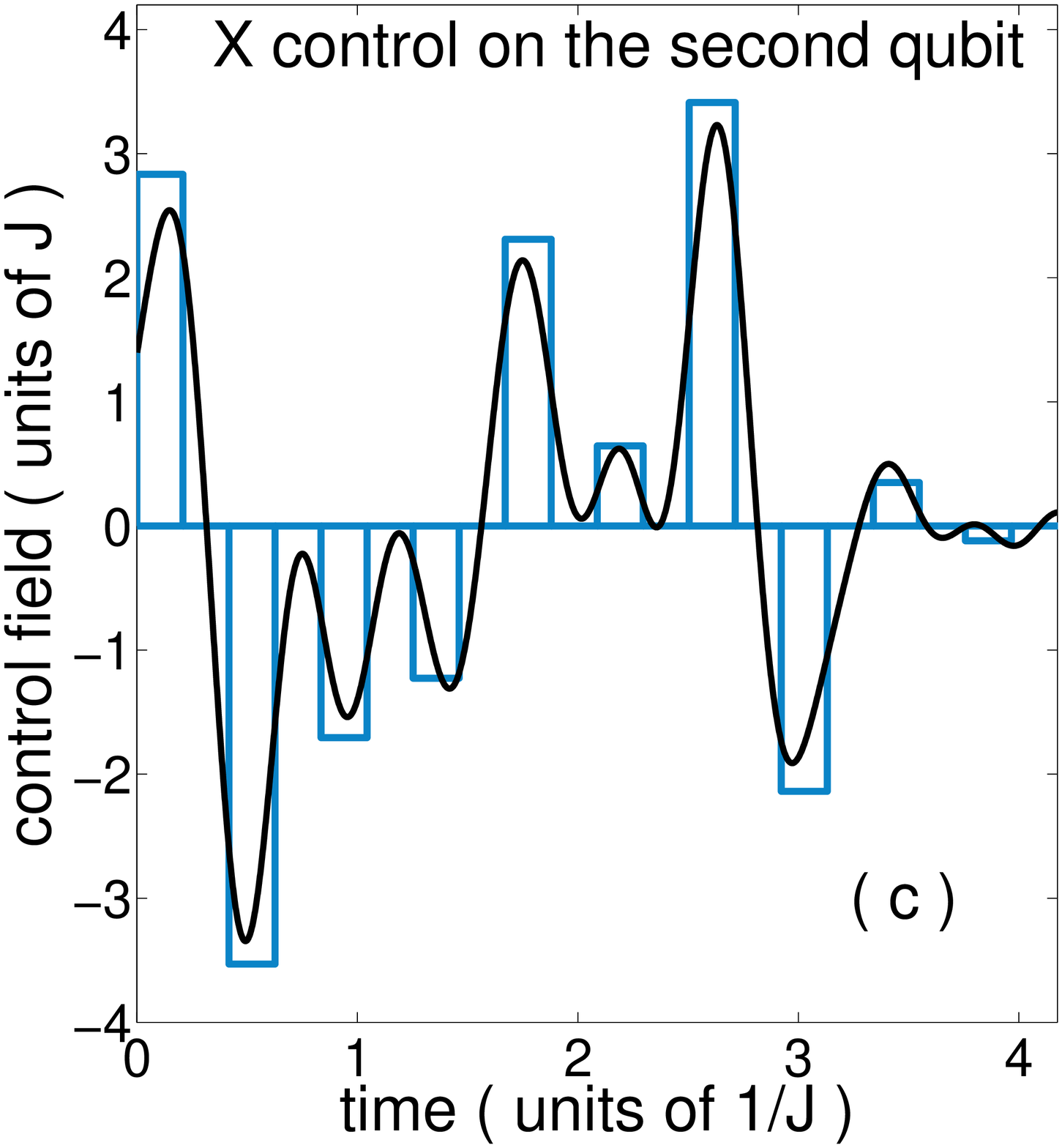}%
\includegraphics[width = 0.425\linewidth,clip=true]{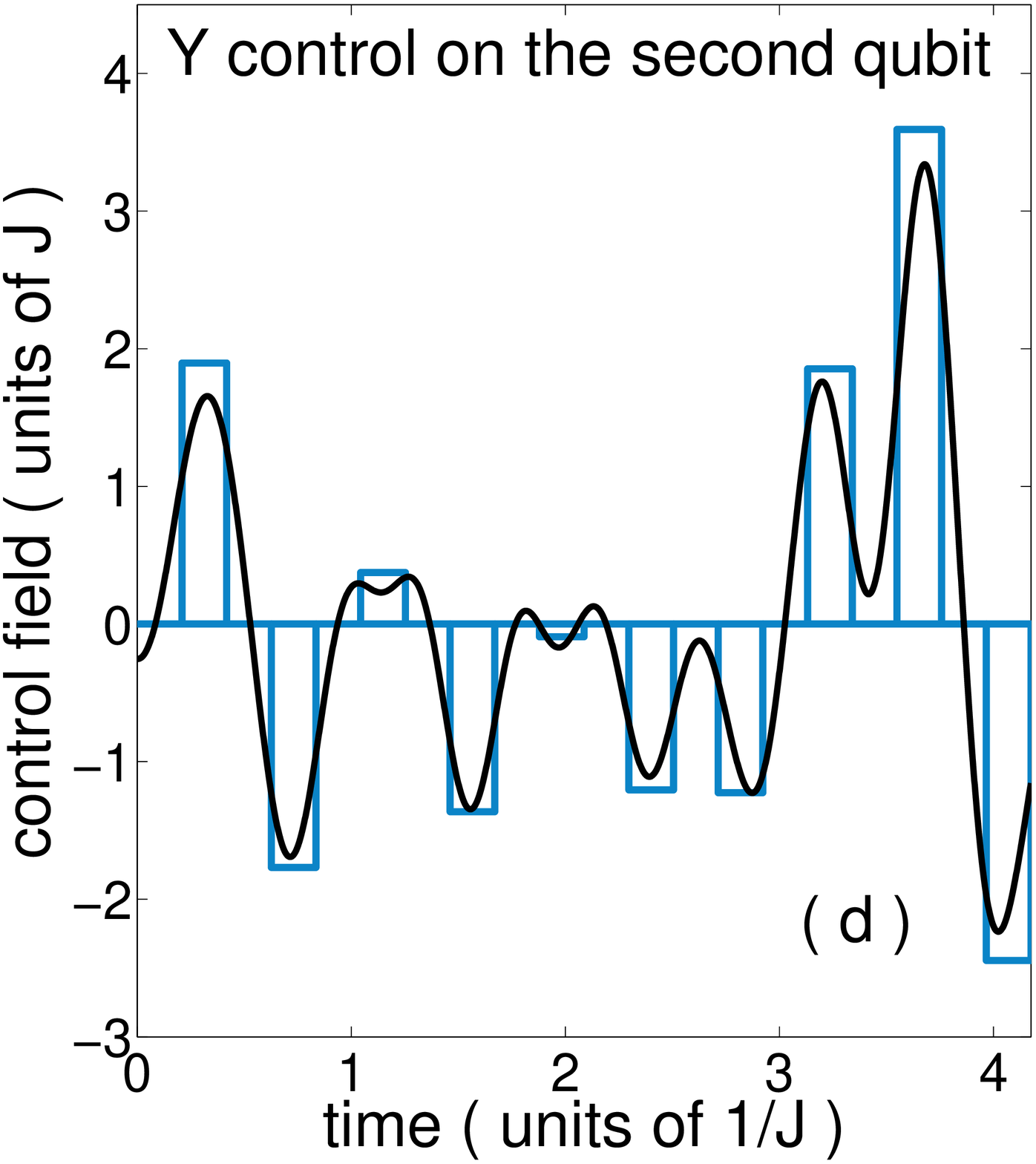}\\
\includegraphics[width = 0.425\linewidth,clip=true]{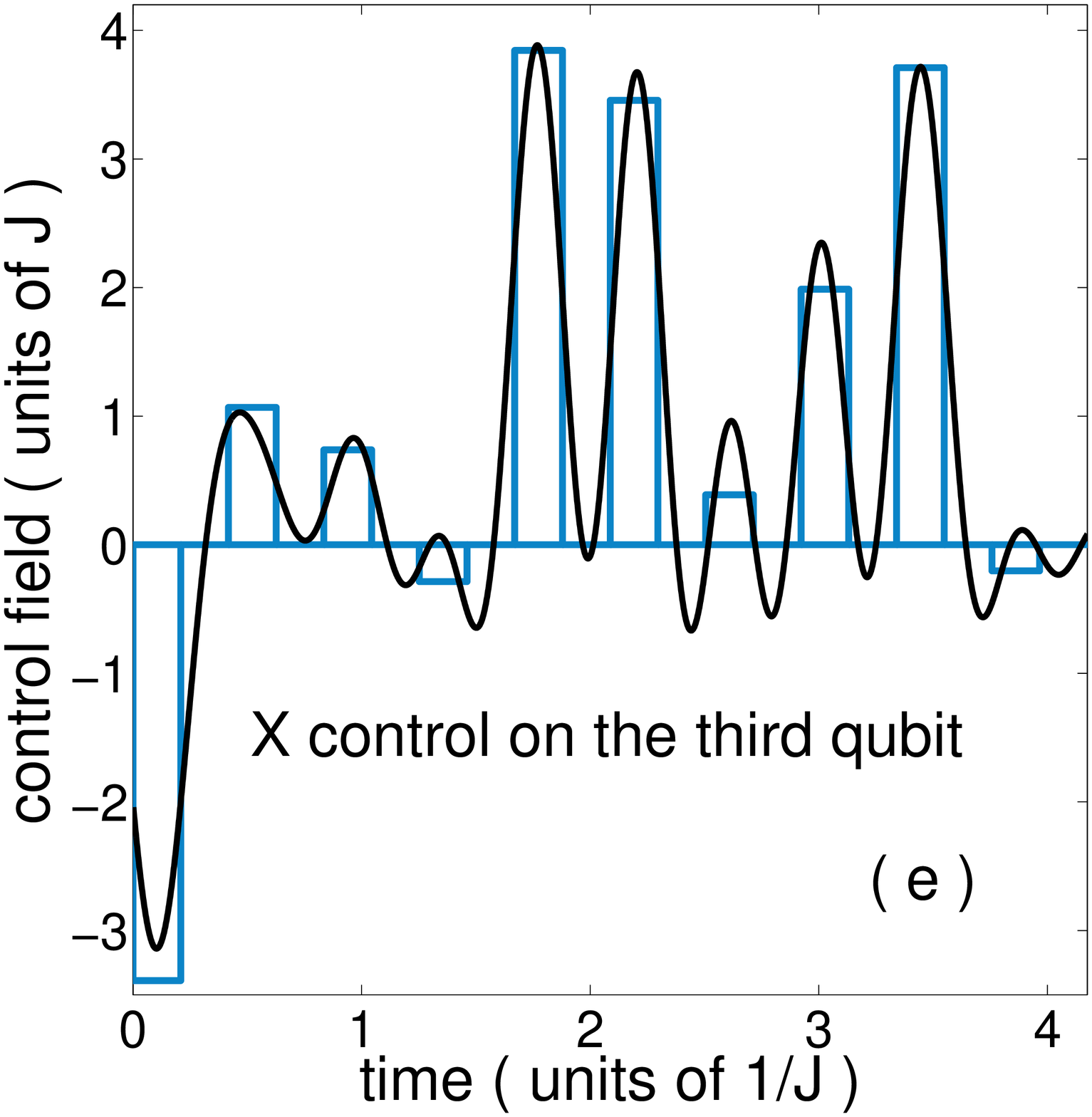}%
\includegraphics[width = 0.425\linewidth,clip=true]{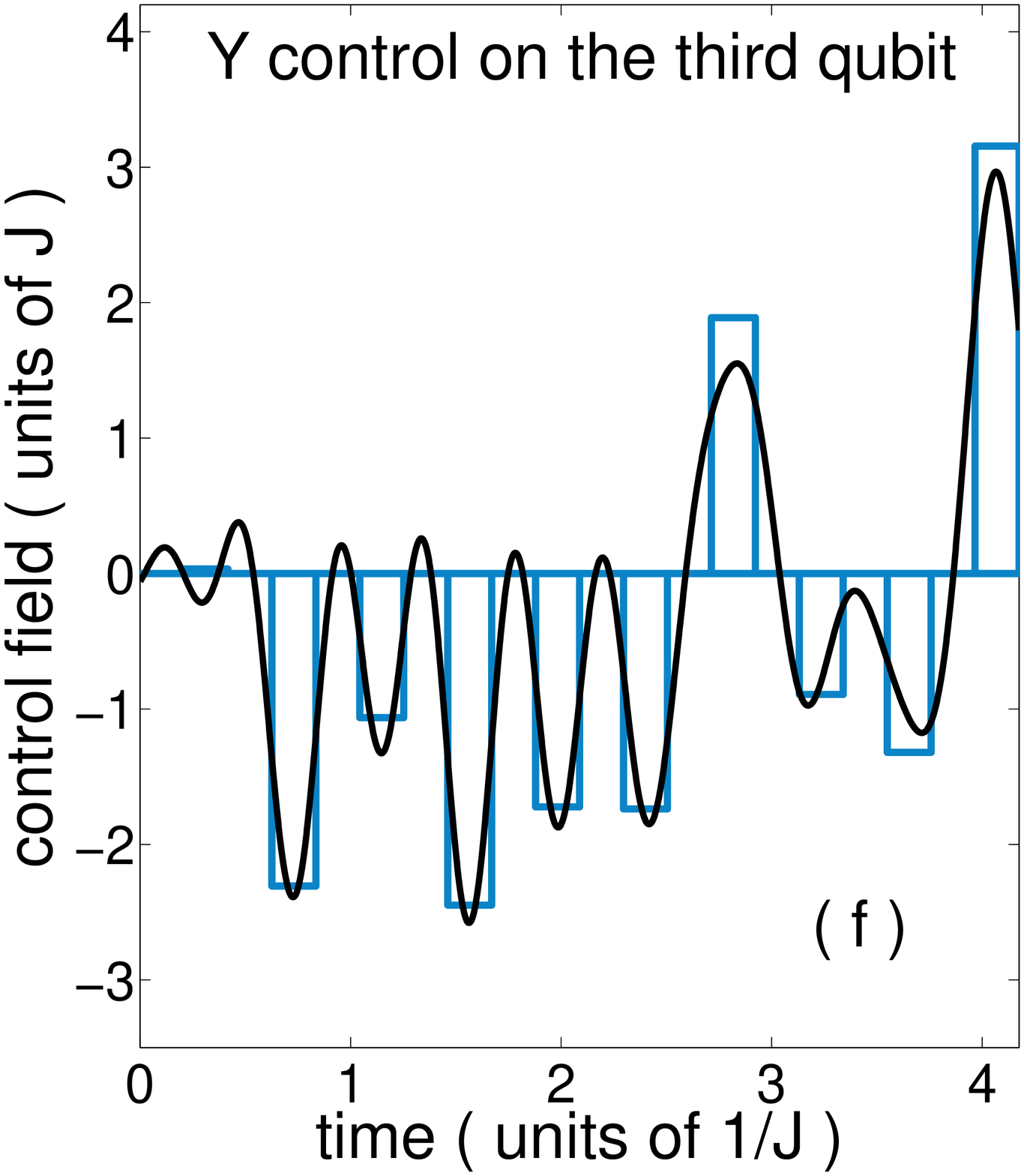}
\caption{\label{optimal_fields} 
Piecewise-constant and filtered ($\omega_{0}=500$\:MHz) 
control fields acting on the three qubits for $J= 30$\:MHz, i.e., a
gate time $t_{g}$ of $4.18 J^{-1} =139$\:ns.}
\end{figure}

To make contact with possible experiments, we now assume $J=30$\:MHz
and $J_{12}=J_{23}=6 J_{13}=J$~\cite{Baur++:11}. 
As can be inferred from the table, a Toffoli gate can be realized in
$2.25 J^{-1}=75$\:ns 
with a fidelity higher than $90\%$, while fidelities larger than $99\%$  
can be reached for gate times of around 
$4.18 J^{-1} = 140$\:ns. 
Examples of optimal $x$ and $y$ control fields on all three qubits 
are shown in Fig.~\ref{optimal_fields}.

For fixed total time $t_{g}=N_{t}T$ higher fidelities are obtained for
larger $N_{t}$ (i.e., smaller $T$), and the same is true of the robustness of 
these fidelities to random errors in the control-field amplitudes~\cite{Heule++}. 
However, for $T$ smaller than some (nonuniversal) threshold value, 
it becomes impossible to reach high fidelities ($F>90\%$) without violating 
the constraint $\widetilde{\Omega}_{\textrm{max}}<130$\:MHz.

In reality, the fidelity loss resulting from decoherence is
inextricably linked to the particular experimental setup and noise
sources present in it. The errors due to decoherence certainly depend
sensitively on the total gate time, which is minimized in our approach
by the interplay of always-on interactions between the qubits and
time-dependent control pulses acting on all qubits. Quite generally,
the suppression of the gate fidelity due to decoherence is
approximately given by the factor $\exp(-t_{g}/T_{2})$, determined by
the ratio of the gate time $t_{g}$ and the decoherence time
$T_{2}$~\cite{DiCarlo++:09}. It is therefore quite encouraging that
the required times we find for high-fidelity ($F>90\%$) realizations
of the Toffoli gate ($t_{g}\sim 75$\:ns) represent a rather small 
fraction of the newly achieved decoherence times ($T_{2}\sim
5-20$\:$\mu$s). Remarkably, this is better than achieved experimentally 
(with $T_{2}\sim\:1$\:$\mu$s) for two-qubit gates ($t_{g}\sim 30-60$\:ns).

To summarize, employing methods of quantum operator control we have
investigated the feasibility of realizing a quantum Toffoli gate with
superconducting qubits in a circuit QED setup. Our calculations
indicate that within $75$\:ns a Toffoli gate can be realized with
intrinsic fidelities higher than $90\%$, while fidelities larger than
$99\%$ require gate times of about $140$\:ns.
A particularly appealing feature of our approach is that it does not
make a principal difference between two- and three qubit gates, in
contrast to the more conventional approaches in which a Toffoli gate
is realized through several two-qubit and single-qubit gates. Our
method can be used for realizing other three-qubit gates (e.g., the
Fredkin gate) and can also be straightforwardly generalized to an
arbitrary number of qubits.

Our study can be extended using more sophisticated iterative schemes
for finding optimal control pulses, and could form the basis of an
open-loop type approach taking into account higher qubit levels and
experimental uncertainties.  The reduced gate times are likely to
simplify the realization of three-qubit Toffoli gates and lead to a
higher fidelity than the direct
approach~\cite{Mariantoni++:11,Fedorov++:11,Reed++:11}.

\begin{acknowledgments}
We acknowledge useful discussions with S. Aldana, S. Filipp, R. Heule,
and A. Nunnenkamp.  This work was financially supported by EU project
SOLID, the Swiss NSF, the NCCR Nanoscience, and the NCCR Quantum
Science and Technology.
\end{acknowledgments}


\begin{thebibliography}{28}
\expandafter\ifx\csname natexlab\endcsname\relax\def\natexlab#1{#1}\fi
\expandafter\ifx\csname bibnamefont\endcsname\relax
  \def\bibnamefont#1{#1}\fi
\expandafter\ifx\csname bibfnamefont\endcsname\relax
  \def\bibfnamefont#1{#1}\fi
\expandafter\ifx\csname citenamefont\endcsname\relax
  \def\citenamefont#1{#1}\fi
\expandafter\ifx\csname url\endcsname\relax
  \def\url#1{\texttt{#1}}\fi
\expandafter\ifx\csname urlprefix\endcsname\relax\def\urlprefix{URL }\fi
\providecommand{\bibinfo}[2]{#2}
\providecommand{\eprint}[2][]{\url{#2}}

\bibitem[{SCq({\natexlab{a}})}]{SCqubitReviews}
\bibinfo{note}{For a review, see Y. Makhlin, G. Sch\"{o}n, and A. Shnirman,
  Rev. Mod. Phys. ${\mathbf{73}}$, 357 (2001); J. Q. You and F. Nori, Phys.
  Today ${\mathbf{58}}$, 42 (2005); J. Clarke and F. K. Wilhelm, Nature
  (London) ${\mathbf{453}}$, 1031 (2008).}

\bibitem[{\citenamefont{Martinis et~al.}(1985)\citenamefont{Martinis, Devoret,
  and Clarke}}]{Martinis+:85}
\bibinfo{author}{\bibfnamefont{J.~M.} \bibnamefont{Martinis}},
  \bibinfo{author}{\bibfnamefont{M.~H.} \bibnamefont{Devoret}},
  \bibnamefont{and} \bibinfo{author}{\bibfnamefont{J.}~\bibnamefont{Clarke}},
  \bibinfo{journal}{Phys. Rev. Lett.} \textbf{\bibinfo{volume}{55}},
  \bibinfo{pages}{1543} (\bibinfo{year}{1985}).

\bibitem[{Bla()}]{Blais++:04}
\bibinfo{note}{A. Blais {\em et al.}, Phys. Rev. A {\bf 69}, 062320 (2004).}

\bibitem[{Wal()}]{Wallraff++:04}
\bibinfo{note}{A. Wallraff {\em et al.}, Nature (London) {\bf 431}, 162
  (2004).}

\bibitem[{Maj()}]{Majer++:07}
\bibinfo{note}{J. Majer {\em et al.}, Nature (London) {\bf 449}, 443 (2007).}

\bibitem[{Koc()}]{Koch++:07}
\bibinfo{note}{J. Koch {\em et al.}, Phys. Rev. A ${\mathbf{76}}$, 042319
  (2007).}

\bibitem[{Kim()}]{Kim++:11}
\bibinfo{note}{Z. Kim {\em et al.}, Phys. Rev. Lett. ${\mathbf{106}}$, 120501
  (2011).}

\bibitem[{Pai()}]{Paik++:11}
\bibinfo{note}{H. Paik {\em et al.}, Phys. Rev. Lett. \textbf{107}, 240501 (2011).}

\bibitem[{DiC()}]{DiCarlo++:09}
\bibinfo{note}{L. DiCarlo {\em et al.}, Nature (London) ${\mathbf{460}}$, 240
  (2009).}

\bibitem[{Mer()}]{MerminBook}
\bibinfo{note}{N. D. Mermin, {\em Quantum Computer Science: An Introduction}
  (Cambridge University Press, New York, 2007).}

\bibitem[{Mon()}]{Monz++:09}
\bibinfo{note}{T. Monz {\em et al.}, Phys. Rev. Lett. ${\mathbf{102}}$, 040501
  (2009).}

\bibitem[{Lan()}]{Lanyon++:09}
\bibinfo{note}{B. P. Lanyon {\em et al.}, Nat. Phys. ${\mathbf{5}}$, 134
  (2009).}

\bibitem[{Fed()}]{Fedorov++:11}
\bibinfo{note}{A. Fedorov {\em et al.}, arXiv:1108.3966, 
Nature \textbf{481}, 170 (2012).}

\bibitem[{Mar()}]{Mariantoni++:11}
\bibinfo{note}{M. Mariantoni {\em et al.}, arXiv:1109.3743, 
Science \textbf{334}, 61 (2011).}

\bibitem[{Ree()}]{Reed++:11}
\bibinfo{note}{M. D. Reed {\em et al.}, arXiv:1109.4948.}

\bibitem[{\citenamefont{D'Alessandro}(2008)}]{D'AlessandroBook}
\bibinfo{author}{\bibfnamefont{D.}~\bibnamefont{D'Alessandro}},
  \emph{\bibinfo{title}{Introduction to {Q}uantum {C}ontrol and {D}ynamics}}
  (\bibinfo{publisher}{Taylor \& Francis}, \bibinfo{address}{Boca Raton},
  \bibinfo{year}{2008}).

\bibitem[{SCq({\natexlab{b}})}]{SCqubitControl}
\bibinfo{note}{S. Montangero, T. Calarco, and R. Fazio, Phys. Rev. Lett.
  ${\mathbf{99}}$, 170501 (2007); M. Steffen, J. M. Martinis, and I. L. Chuang,
  Phys. Rev. B ${\mathbf{68}}$, 224518 (2003); P. Rebentrost and F. K. Wilhelm,
  {\em ibid.} ${\mathbf{79}}$, 060507(R) (2009); R. Fisher {\em et al.}, {\em
  ibid.} ${\mathbf{81}}$, 085328 (2010); H. Yuan {\em et al.}, Phys. Rev. A
  ${\mathbf{79}}$, 042309 (2009); J. M. Chow {\em et al.}, {\em ibid.}
  ${\mathbf{82}}$, 040305 (2010).}

\bibitem[{Mot()}]{Motzoi++:09}
\bibinfo{note}{F. Motzoi {\em et al.}, Phys. Rev. Lett. ${\mathbf{103}}$,
  110501 (2009).}

\bibitem[{Gam()}]{Gambetta++:11}
\bibinfo{note}{J. M. Gambetta {\em et al.}, Phys. Rev. A ${\mathbf{83}}$,
  012308 (2011).}

\bibitem[{Gal()}]{GaliautdinovGeller}
\bibinfo{note}{A. Galiautdinov, Phys. Rev. A ${\mathbf{75}}$, 052303 (2007); M.
  R. Geller {\em et al.}, {\em ibid.} ${\mathbf{81}}$, 012320 (2010); J. Ghosh
  and M. R. Geller, {\em ibid.} ${\mathbf{81}}$, 052340 (2010).}

\bibitem[{\citenamefont{Zhang and Whaley}(2005)}]{Zhang+Whaley:05}
\bibinfo{author}{\bibfnamefont{J.}~\bibnamefont{Zhang}} \bibnamefont{and}
  \bibinfo{author}{\bibfnamefont{B.}~\bibnamefont{Whaley}},
  \bibinfo{journal}{Phys. Rev. A} \textbf{\bibinfo{volume}{71}},
  \bibinfo{pages}{052317} (\bibinfo{year}{2005}).

\bibitem[{Loc()}]{LocalControl}
\bibinfo{note}{S. G. Schirmer, I. C. H. Pullen, and P. J. Pemberton-Ross, Phys.
  Rev. A ${\mathbf{78}}$, 062339 (2008); D. Burgarth {\em et al.}, {\em ibid.}
  ${\mathbf{79}}$, 060305(R) (2009).}

\bibitem[{Heu()}]{Heule++}
\bibinfo{note}{R. Heule {\em et al.}, Phys. Rev. A ${\mathbf{82}}$, 052333
  (2010); Eur. Phys. J. D ${\mathbf{63}}$, 41 (2011).}

\bibitem[{Fil()}]{Filipp++:11}
\bibinfo{note}{S. Filipp {\em et al.}, Phys. Rev. A {\bf 83}, 063827 (2011).}

\bibitem[{Bau()}]{Baur++:11}
\bibinfo{note}{M. Baur {\em et al.}, arXiv:1107.4774.}

\bibitem[{\citenamefont{Fazio et~al.}(1999)\citenamefont{Fazio, Palma, and
  Siewert}}]{Fazio+:99}
\bibinfo{author}{\bibfnamefont{R.}~\bibnamefont{Fazio}},
  \bibinfo{author}{\bibfnamefont{G.~M.} \bibnamefont{Palma}}, \bibnamefont{and}
  \bibinfo{author}{\bibfnamefont{J.}~\bibnamefont{Siewert}},
  \bibinfo{journal}{Phys. Rev. Lett.} \textbf{\bibinfo{volume}{83}},
  \bibinfo{pages}{5385} (\bibinfo{year}{1999}).

\bibitem[{NRf()}]{NRfortranBook}
\bibinfo{note}{W. H. Press {\em et al.}, {\em Numerical Recipes in Fortran 77
  and 90: The Art of Scientific and Parallel Computing} (Cambridge University
  Press, Cambridge, 1997).}

\bibitem[{Mac()}]{Machnes++:11}
\bibinfo{note}{See, e.g., S. Machnes {\em et al.}, Phys. Rev. A
  ${\mathbf{84}}$, 022305 (2011).}


\end{thebibliography}

\end{document}